# Ideal Weak Topological Insulator and Protected Helical Saddle Points

Ji Seop Oh[1,2], Tianyi Xu[3], Nikhil Dhale[3], Sheng Li[3], Chao Lei[4], Chiho Yoon[3], Wenhao Liu[3], Jianwei Huang[2], Hanlin Wu[3], Makoto Hashimoto[5], Donghui Lu[5], Chris Jozwiak[6], Aaron Bostwick[6], Eli Rotenberg[6], Chun Ning Lau[7], Bing Lv[3,\*], Fan Zhang[3,\*], Robert Birgeneau[1,8,\*], Ming Yi[2,\*]

[1]Department of Physics, University of California, Berkeley, California 94720, USA

[2]Department of Physics and Astronomy, Rice University, Houston, Texas 77024, USA

[3]Department of Physics, The University of Texas at Dallas, Richardson, Texas 75080, USA

[4]Department of Physics, The University of Texas at Austin, Austin, Texas 78712, USA

[5]Stanford Synchrotron Radiation Lightsource, SLAC National Accelerator Laboratory, Menlo Park, California 94025, USA

[6]Advanced Light Source, Lawrence Berkeley National Laboratory, Berkeley, California 94720, USA

[7]Department of Physics, The Ohio State University, Columbus, Ohio 43210, USA

[8]Materials Science Division, Lawrence Berkeley National Laboratory, Berkeley, California 94720, USA

**The paradigm of classifying three-dimensional (3D) topological insulators into strong and weak ones (STI and WTI) opens the door for the discovery of various topological phases of matter protected by different symmetries and defined in different dimensions. However, in contrast to the vast realization of STIs, very few materials have been experimentally identified as being close to WTI. Even amongst those identified, none exists with topological surface states (TSS) exposed in a global bulk band gap that is stable at all temperatures. Here we report the design and observation of an ideal WTI in a quasi-one-dimensional (quasi-1D) bismuth halide, $Bi_4I_{1.2}Br_{2.8}$ (BIB). Via angle-resolved photoemission spectroscopy (ARPES), we identify that BIB hosts TSS on the (100)' side surface in the form of two anisotropic $\pi$-offset Dirac cones (DCs) separated in momentum while topologically dark on the (001) top surface. The ARPES data fully determine a unique side-surface Hamiltonian and thereby identify two pairs of non-degenerate helical saddle points and a series of four Lifshitz transitions. The fact that both the surface Dirac and saddle points are in the global bulk band gap of 195 meV, combined with the small Dirac velocities, nontrivial spin texture, and the near-gap chemical potential, qualifies BIB to be not only an ideal WTI but also a fertile ground for topological many-body physics.**

In 3D bulk crystals, while a STI hosts an odd number of gapless Dirac surface states on every surface, a WTI is characterized by an even number of such surface states on selected surfaces[1,2]. The distinction between the labels "weak" versus "strong" originates from a perception that the TSS of a WTI could be easily gapped via annihilation of the even number of DCs with $Z_2$ character in the presence of translation symmetry breaking, e.g., by disorder, thereby making the WTI indistinguishable from a trivial insulator whereas a STI is robust[3]. Recent theoretical developments have shown, however, that not only is the gapless nature of WTIs robust to disorder[4], but also the initially perceived vulnerability bestows upon WTIs a desired degree of tunability: layer-dimerization gapping TSS in turn gives birth to gapless hinge modes—a mechanism for a topological phase transition into a higher-order TI[5–7]. Hence, ideal WTIs possess much stronger potential for tunable transitions between 3D topologically distinct phases of matter and, as elucidated below, for realizing two-dimensional (2D) topological many-body physics.

Compared to the vast number of STIs experimentally identified, however, WTIs are fundamentally challenging to realize experimentally, as implied by the two theoretically proposed pathways. In the first pathway, WTIs are found as periodic stacks of weakly coupled 2D quantum spin Hall insulators, and for these layered materials the TSS are located on the non-cleavable side surfaces[8]. The second is to engineer a superlattice of alternating layers with multiple band inversions that requires fine tuning of intra- and inter-layer couplings[9,10]. Besides these challenges, the TSS in all the WTIs reported so far, such as ZrTe$_5$[11] and RhBi$_2$[12], are buried deep in bulk states. Recently, a new series of quasi-1D material bismuth halides Bi$_4$X$_4$ (X = I, Br) has emerged[6,7,13–16]. Consisting of weakly-coupled quasi-1D Bi-X chains, the bismuth halides provide two different naturally cleavable surfaces accessible for ARPES characterization of TSS[7,14]. Amongst these, $\beta$-Bi$_4$I$_4$ has been identified as a WTI that is only stable above 300 K, below which a structural dimerization gaps out the TSS, leading to a higher-order TI[7]. The high temperature range in which the WTI phase is stable makes it challenging to characterize the unique TSS (not to mention the utilization in quantum devices), leading to controversial ARPES observations and interpretations[7,13,14].

In order to realize a material platform with stable, tunable, and accessible surface-selective TSS, we designed a WTI in the form of Bi$_4$I$_{1.2}$Br$_{2.8}$ (BIB). Guided by the experimental confirmation that the $\beta$-phase is stable in Bi$_4$I$_4$ but absent in Bi$_4$Br$_4$[17–19] and by the theoretical prediction that, if synthesized, $\beta$-Bi$_4$Br$_4$ is a WTI superior to $\beta$-Bi$_4$I$_4$ because of a larger bulk gap[16], we searched for an optimized substitution ratio between I and Br for a single-stack crystal structure that is stable at all temperatures down to 0 K. We realized this optimization at the ratio of I:Br = 1.2:2.8 and its vicinity (see Methods). The basic building block of the quasi-1D crystal structure of BIB is the Bi-X chain along the *b* axis (Fig. 1a). The chain stacks along both the *a* and *c* axes, leading to two distinct natural cleavage planes—the (001) top surface exposing the *ab* plane and the (20$\bar{1}$) side surface exposing the plane parallel to ***b*** and ***a*** + 2***c***. Single crystal X-ray diffraction demonstrates that BIB has only one single structural phase with a unit cell containing only one single layer of chains, which is reminiscent of the high temperature $\beta$- phase of Bi$_4$I$_4$. The (20$\bar{1}$) direction in BIB corresponds to (100) in $\beta$-Bi$_4$I$_4$, where the difference originates from the stacking structure. For simplicity we will refer to it as (100)' surface hereafter (See Supplementary Information for details). Temperature-dependent resistivity curves also confirm the lack of a structural distortion down to 2 K (Fig. 1b). Large single crystals of BIB with extended thickness along both the *a* and *c* axes have been grown (Fig. 1c), allowing direct accesses to the (001) and the (100)' surfaces in ARPES measurements without the need for a nano-focused beam spot.

Next, we examine the experimental accessibility to the two inequivalent surfaces via ARPES. Contrary to STIs manifesting an odd number of gapless DCs on every surface, WTIs host an even number of DCs only on selected surfaces[3,16]. Therefore, the absence/presence of non-trivial surface states for each surface can provide definitive evidence for experimental identification of the topological character of a potential WTI candidate. Figures 1e and f show overviews of momentum-energy dispersions from the (001) and (100)' surfaces, respectively. We observe distinct electronic states, signifying unambiguous single-facet observations for each measurement under a 40 μm beam spot. In particular, the Fermi surface (FS) map from the (001) surface displays a spot-like feature at the $\bar{M}$ point (Fig. 1e), in sharp contrast to the parallel line FS across the whole (100)'-projected BZ (Fig. 1f). The clear difference in FS reveals that the interlayer interaction along the *a* axis is stronger than that along the *c* axis, consistent with the observations for Bi$_4$I$_4$[7,14]. This is corroborated by the more dispersive bands on the (001) surface (Fig. 1g) than that on the (100)' surface (Fig. 1h). Moreover, the periodicity of the measured dispersions perpendicular to the chains from each surface is inversely related to the corresponding lattice constant, i.e., 1.0 Å$^{-1}$ on the (001) surface and 0.6 Å$^{-1}$ on the (100)' surface (Figs. 1g and h), and hence can be used to confirm unambiguously the surface termination (Fig. 1d).

Having demonstrated the ability to distinguish the two measured surfaces, we classify the topological character of each surface (Fig. 2). First from the (001) surface, we investigate band dispersions along the chain at the two time-reversal-invariant momenta (TRIM) $\bar{\Gamma}$ and $\bar{M}$. While no band crossing is observed at $\bar{\Gamma}$ (Fig. 2d), the high symmetry cut across $\bar{M}$ shows two spectral features visible in the spectral image in Fig. 2f—a valence band with a band top at -0.24 eV and a conduction band with a band bottom just below $E_F$. These bands are also guided by the dotted grey lines in the energy distribution curves (EDCs) in Fig. 2g. To obtain the precise band gap at $\bar{\text{M}}$, we fit the EDC at $k_y = 0$ (the red curve in Fig. 2g) with three Lorentzian peaks multiplied by the Fermi-Dirac distribution convolved with Gaussian broadening from the experimental resolution (Fig. 2h). The observed variation of the band gap with photon energies reveals that the gap is of bulk nature, varying from 195 meV when measured with 126 eV photons to 210 meV with 118 eV photons. Thus, we identify the minimum gap size of 195±10 meV to be the global bulk gap for BIB. We note that the bulk band gap of $\beta$-Bi$_4$I$_4$ is smaller than 100 meV[7], less than what we find in BIB. To summarize, we do not observe any in-gap states on the (001) surface, demonstrating the absence of any topologically non-trivial (001) surface state.

By contrast, the (100)' side surface exhibits surface-originated linear crossings at two different TRIM. In particular, at both $\bar{\Gamma}$ (Fig. 2i) and $\bar{\text{Z}}$ (Fig. 2l), the band dispersions show simple linear crossings slightly below $E_F$. In a zoomed-in view, the crossing at $\bar{\Gamma}$ appears at -110 meV (Fig. 2j) while that at $\bar{\text{Z}}$ appears at -140 meV (Fig. 2m). The energy positions of these crossings do not vary with the photon energy, indicating their surface nature (Fig. S6). To demonstrate the gapless nature of both crossings at $\bar{\Gamma}$ and $\bar{\text{Z}}$, we plot the EDCs in Figs. 2k and n. Guides on the band dispersions are plotted in grey and cyan to indicate respectively the bulk and surface states as determined by the photon energy dependence (Supplementary Information). Noticeably, the EDCs at $k_y = 0$ for both TRIM have a single peak that splits into two peaks away from $k_y = 0$. To demonstrate this, we fit the EDCs at $k_y = 0$ and $k_y = \pm 0.01$ Å$^{-1}$ using Lorentzian functions with Fermi-Dirac distribution convolved with the experimental resolution (Figs. 2o and p). The two peaks plotted in grey at high binding energies are bulk states. To obtain reliable fitting results for the low-energy surface state spectra, two peaks are necessary for EDCs at $k_y = \pm 0.01$ Å$^{-1}$ while only one is needed for $\bar{\Gamma}$ and $\bar{\text{Z}}$, demonstrating that the crossings are gapless. Thus, we confirm that the crossings from the (100)' side surface are a pair of gapless Dirac surface states.

Taking the observations of the topologically dark (001) surface and two gapless Dirac surface states on the (100)' surface together, we have demonstrated the necessary conditions to identify BIB as a WTI. To verify this theoretically, we performed first-principles calculations for BIB (Methods). As shown in Fig. 2c, the bulk electronic states of BIB near the Fermi level are composed of Bi 6*p* states. In particular, two band inversions between the interior and exterior Bi $p_x$ (see Fig. 1a) bands occur at the L and M points in the 3D BZ, which would annihilate when projected onto the (001) surface but be preserved as two linear Dirac crossings when projected onto the (100)' surface. Thus, BIB is a WTI with $Z_2$ invariants (0;001) based on the Fu-Kane criterion[20] (see Supplementary Information for details).

The existence of two distinct DCs on the (100)' surface, together with the quasi-1D geometry, leads to a unique topological structure with two protected pairs of helical saddle points. To illustrate this, we construct an effective Hamiltonian[6] constrained by the quasi-1D geometry, surface symmetry, and WTI topology for the experimentally discovered (100)' surface states of BIB. BIB has an identical space group with that of *β*-Bi$_4$I$_4$ (C12/*m*1, No. 12), and its (100)' surface has only the (010) mirror and time-reversal symmetries. Thus, the minimal effective Hamiltonian can be written as (see Methods)

$$H = \frac{t}{2}[\cos(k_c c + k_1 a_1) + \cos(k_c c + k_2 a_2)] + v_y \hbar k_y \sigma_z + v_z \hbar[\sin(k_c c + k_1 a_1) + \sin(k_c c + k_2 a_2)]/d\, \sigma_y$$

where $2t = 30$ meV is the energy difference between two Dirac points, $d = |2\boldsymbol{c} + \boldsymbol{a}| = 20.335$ Å is the conventional lattice constant, $c = 10.456$ Å is the primitive lattice constant in the stacking direction, and $v_{y,z}$ are the Dirac velocities to be determined in the chain and stacking directions, respectively. The surface state structure of this minimal model is shown in Fig. 3a, where there are two pairs of saddle points. The presence of these surface saddle points is a direct consequence of the weak dispersion in the stacking direction and the ultimate merger of the two surface DCs. The maximum and minimum of the sinusoidal dispersion correspond to the upper and lower pair of saddle points. Distinct from those in other scenarios, these non-degenerate saddle points inherit the helical nature from the DCs. Moreover, we can quantitatively determine the values of $v_{y,z}$ from a comparison of the topological characteristics between the model predictions and the experimental observations. We determine $v_y = 3.3 \times 10^6$ m/s from the linear band dispersions at $\bar{\Gamma}$ and $\bar{Z}$ and $v_z = 1.9 \times 10^4$ m/s from a direct comparison of the theoretically calculated (Fig. 3d) and experimentally obtained (Fig. 3e) constant energy surfaces (CESs). Note that this model is constructed from the experimentally determined crystal structure and informed by the experimentally observed gapless DCs at $\bar{\Gamma}$ and $\bar{Z}$. The presence of the pairs of helical saddle points and the energy gap between $\bar{\Gamma}$ and $\bar{Z}$ are guaranteed, because the two DCs must be connected between $\bar{\Gamma}$ and $\bar{Z}$, and because time-reversal symmetry cannot protect any extra crossing between $\bar{\Gamma}$ and $\bar{Z}$.

Having fully constrained the minimal model, we can now precisely determine the location of the helical saddle points, which are labeled along with the Dirac points on a display of the measured dispersions (Figs. 3a and c). We also overlay the bulk conduction band (BCB) and bulk valence band (BVB) locations determined previously in Fig. 2h and highlighted in yellow. From this visualization, the entire surface state structure displayed in Fig. 3c is located within the global bulk band gap, including the two surface DCs and the two pairs of helical saddle points. This justifies that BIB is an unambiguous WTI that exhibits the ideal WTI surface characteristics. The energy and momentum differences between a saddle point and its nearest Dirac point are $\sqrt{t^2 + 4v_z^2\hbar^2/d^2} - t$ and $2\tan^{-1}(2v_z\hbar/td)/d$, which are 4.4 meV and 0.07 Å$^{-1}$, respectively. Thus, it is challenging to directly resolve the saddle points separately from the Dirac points, given that the energy resolution of our ARPES is ~20 meV.

Next, we examine the evolution of the CESs as a function of energy as each of the aforementioned WTI surface characteristics is crossed; there is a series of four Lifshitz transitions[16]. Starting from 50 meV below $E_F$, the two-line electron-like CES is open and crosses the entire BZ. Moving down in energy, the open CES shrinks and is pinched into two closed electron pockets by the higher-energy helical saddle points (SP$_1$), indicating the first Lifshitz transition. Going further down in energy and crossing the higher-energy Dirac point (DP$_1$), the second Lifshitz transition develops: the electron pocket centered at $\bar{\Gamma}$ first contracts to a point at DP$_1$ and then expands into a hole pocket below DP$_1$. Below DP$_1$ and above the lower-energy Dirac point (DP$_2$), the electron pocket at $\bar{\Gamma}$ and the hole pocket at $\bar{Z}$ coexist. When reaching DP$_2$, the third Lifshitz transition occurs with the $\bar{Z}$ pocket switching from electron-like to hole-like via DP$_2$. Ultimately, the two hole pockets merge into the two-line open hole-like CES when the lower-energy helical saddle points (SP$_2$) is crossed, leading to the fourth Lifshitz transition.

A few remarks are in order. The existence of two pairs of helical saddle points, overlooked before, is a part of the topological structure of the unique WTI surface state: the double DCs are $\pi$-offset in momentum, which is in sharp contrast to STI with a single surface Dirac cone. However, in general, the Dirac and saddle points are not necessarily both in the global bulk gap in a WTI as in $ZrTe_5$ and $RhBi_2$. Here in BIB, it is the weak inter-layer coupling that ensures the Dirac velocity $v_z$ to be small and hence the saddle points to be in-gap, reinforcing BIB to be an ideal WTI. Recently, it has been suggested that one can artificially engineer Dirac flat bands by overlaying two STIs with a small twist[21,22], analogous to twisted bilayer graphene[23]. We emphasize that, in an ideal WTI such as BIB, it is naturally guaranteed to host two distinct DCs in the bulk gap with a narrow bandwidth between the two van Hove singularities (vHSs). In the case of BIB, this bandwidth is ~ 40 meV (Fig. 3c), which can be made even smaller by trimerizing the layers or adding spacer layers, both of which are realizable amongst the class of layered bismuth halides. Last but not least, the non-degenerate surface CESs have non-trivial spin texture and can be closed or open.

The exotic topological manifold observed on the side surface of an ideal WTI could be a fertile ground for topological many-body physics. One example is coupled helical Luttinger-liquid behavior when the Fermi level is tuned beyond either of the two vHSs but still in the bulk gap, i.e., where the corresponding CES is open. Indeed, the weak dispersion along the LM direction (Fig. 2c and Fig. 3) suggests that each (001) monolayer of BIB is a large-gap quantum spin Hall insulator, as confirmed by our first-principles calculations. Appealingly, when such a surface state is proximitized[24] by a linear junction of a ferromagnetic insulator and an *s*-wave superconductor, a Majorana chain with precisely one Majorana zero mode per layer and uniform nearest-neighbor couplings are expected[25]. (Fig. 4a) Another example is unconventional indirect exciton condensation when the Fermi level is tuned to the middle point between the $\bar{\Gamma}$ and $\bar{Z}$ Dirac point energies[26]. In this case, the electron and hole pockets are of the same size and thus enjoy perfect nesting (Fig. 4b). Notably, the spin textures of the two pockets are only half aligned for both the intra- and inter-surface pairing channels, given their helical characteristics revealed by the effective model.

To summarize, we have designed and discovered an ideal WTI in BIB that is topologically dark on the (001) top surface yet hosts two $\pi$-offset DCs and two pairs of helical saddle points on the (100)' side surface. Excitingly, the complete set of topological objects on the (001) surface state is exposed within the global bulk gap and hence accessible to quantum transport. As the chemical potential of bulk crystals is observed here to be at the edge of the bulk gap, gate tuning allows access to the various regimes in which novel physics is expected, given that gate-tunable boundary transport has recently been achieved in $Bi_4I_4$ field effect transistors[27]. Both large critical supercurrent in Josephson junction[27] and superconductivity under hydrostatic pressure[28–30] have recently been discovered in the bismuth halide series, pointing to the potential topological[25]. Altogether, the ideal WTI, BIB, offers an unprecedented surface platform for exploring the interplay between band topology and fermi-surface instabilities.

**Methods**

**Single crystals growth and structural determination.** The single crystals of $Bi_4I_{1.2}Br_{2.8}$ (BIB) were grown using the chemical vapor transport method similar to the growth of $Bi_4I_4$[7]. The Bi (Alfa Aesar 99.999%) pieces, $HgI_2$ (Alfa Aesar 99+%) powder and $HgBr_2$ (Alfa Aesar 99+%) in their respective molar ratios, were mixed and ground in agate mortar, first sealed in an evacuated quartz tube, and then put into a two-zone horizontal tube furnace with a temperature gradient of ~60 °C between hot end (270 °C) and cold end (210°C). After two weeks of reaction, clusters of needle-like crystals were formed with typical size of $3.0 \times 0.5 \times 0.2$ mm$^3$. The whole assembly was further annealed in a low temperature oven at 160 °C for more than 4 weeks to ensure the high quality of the grown crystals. The structure of the phase is determined by single crystal X-ray diffraction using a Bruker Apex DUO X-ray single crystal diffractometer.

**Electrical transport.** Electrical resistivity was measured with a Physical Property Measurement System (Quantum Design) using standard four-probe technique. The current was applied along the *b*-axis of both the $Bi_4I_4$ and BIB crystals.

**Angle-resolved photoemission spectroscopy (ARPES).** ARPES experiments were conducted at the MAESTRO beamline (BL 7.0.2) at the Advanced Light Source and beamline 5-2 at the Stanford Synchrotron Radiation Lightsource with a R4000 electron analyzer with deflector option and a DA30 electron analyzer, respectively. BIB crystals were exfoliated with kapton tape in ultra-high vacuum with a base pressure lower than $5 \times 10^{-11}$ Torr. Energy resolution used was better than 20 meV, as confirmed by measuring a copper puck in vacuum. Angular resolution was 0.1°. Beam spot size was 40 μm × 20 μm, small enough to measure separately the (001) top and (100)' side surfaces. All measurements were carried out below 15 K.

**First-principles calculations.** Density-functional-theory calculations were performed by using the Vienna ab initio simulation package[31] in which the generalized gradient approximations of Perdew-Burke-Ernzerhof have been adopted for exchange-correlation potential[32,33]. During the self-consistent calculations, the global break condition for the electronic SC-loop was set to be $10^{-6}$ eV. The plane-wave-energy cutoff was set to be 300 eV, the Brillouin zone was sampled by a $9 \times 9 \times 6$ mesh, and the Gaussian smearing width was set to be 0.01 eV. Heyd-Scuseria-Ernzerhof (HSE06) hybrid functional method[34] was further employed to improve the electronic band structures, using a 300 eV plane-wave-energy cutoff and a $6 \times 6 \times 4$ Brillouin zone mesh. A threshold of SC-loop was set to be $10^{-6}$ eV in the HSE06 calculations. (These were applied to the study of $Bi_4Br_4$ and of $Bi_4I_4$ before[6,7,16].)

Virtual crystal approximation (VCA)[35] was then employed to simulate the $β$ phase of $Bi_4Br_xI_{4-x}$, in which the Br or I atoms each occupy two Wyckoff positions. At each Wyckoff position there are two atoms connected by the inversion symmetry. In the VCA calculations, while the average $x$ is set to be 2.8, the value at each Wyckoff position varies from 1.6 to 4.0. All these different VCA calculations consistently show that the $x = 2.8$ phase is a weak topological insulator.

**Data availability**

The data presented in this paper will be available from the corresponding authors upon reasonable request.

**Acknowledgements**

This work is mainly supported by National Science Foundation (NSF) through the DMREF program. This research used resources of the Advanced Light Source and the Stanford Synchrotron Radiation Lightsource, both U.S. DOE Office of Science User Facilities under contract Nos. DE-AC02-05CH11231 and DE-AC02-76SF00515, respectively. The work at UC Berkeley is supported by NSF Grant No. DMR-1921798. The work at Rice is supported by NSF under Grant No. DMR-1921847, the Robert A. Welch Foundation under Grant No. C-2024, and the Gordon and Betty Moore Foundation's EPiQS Initiative through grant No. GBMF9470. We acknowledge the Texas Advanced Computing Center (TACC) for providing resources that have contributed to the research results reported in this work. The work at UT Dallas is supported by NSF under Grant Nos. DMR-1921581 and DMR-1945351, AFOSR under Grant No. FA9550-19-1-0037, and Army Research Office (ARO) under Grant No. W911NF-18-1-0416.

**Author contributions**

The project was conceived by MY, RJB, FZ, BL, and CNL. The single crystals were synthesized and characterized by ND, HW, and SL under the guidance of BL. The ARPES measurements were carried out by JSO and JH with the help of MH, DL, CJ, AB, and ER under the guidance of MY and RJB. The first-principle calculations were carried out by CL and theoretical modeling by TX under the guidance of FZ. The manuscript was written by JSO, MY, FZ, and RJB and contributed by all the authors.

**Competing interests**

The authors declare no competing interests.

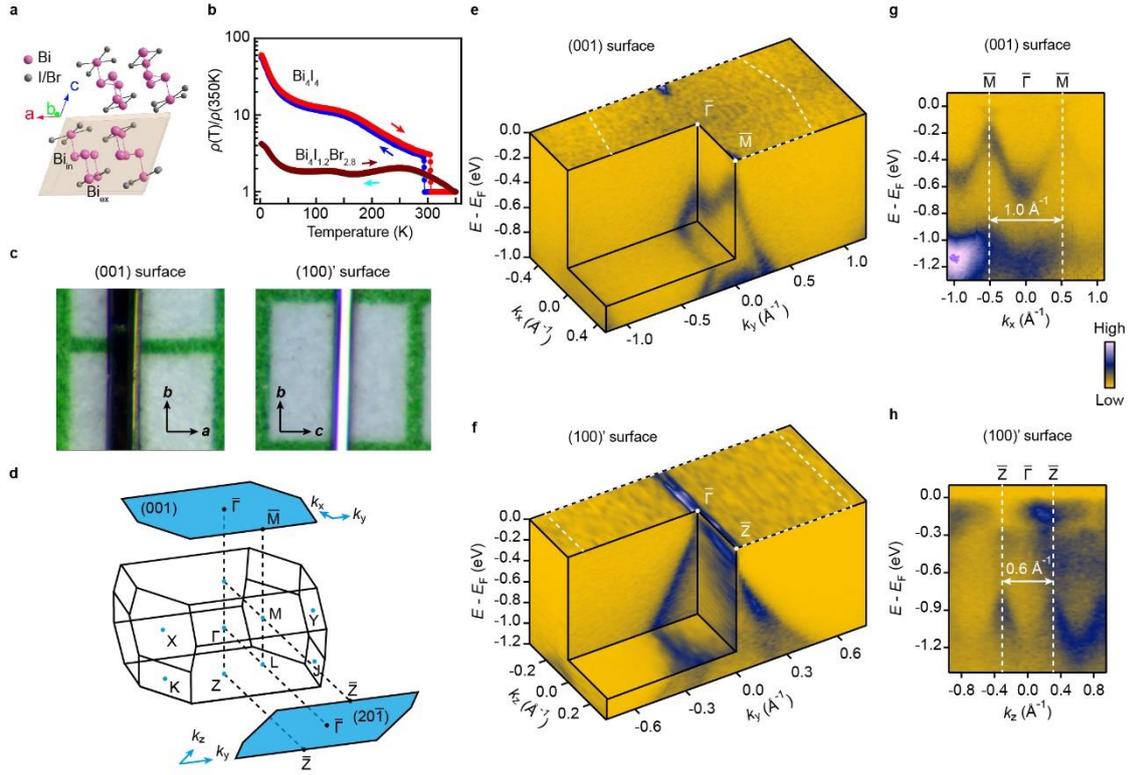

**Fig. 1 Characterizations for BIB and its surface-selective ARPES spectra**

**a,** Crystal structure of $Bi_4I_{1.2}Br_{2.8}$ (BIB), where the Bi-(I, Br) chains are oriented along the *b* axis. A parallelepiped shows the unit cell. **b,** Temperature-dependent resistivity for $Bi_4I_4$ and BIB crystals. The first-order structural transition at 300 K in $Bi_4I_4$ is absent in BIB down to 5 K. **c,** Images for BIB crystals showing the accessible top (001) and side (100)' surfaces. The size of the grid is $1 \times 1$ mm². **d,** Bulk BZ and the (001) and (100)' surface BZs. **e,f,** Visualized ARPES spectra with FS maps on the (001) and (100)' surfaces. 86 eV and 70 eV photons were used for the (001) and (100)' surfaces, respectively. **g,h,** Band dispersions perpendicular to the 1D Bi-(I,Br) chains on the (001) and (100)' surfaces. Periodicities along $k_x$ and $k_z$ are different due to the different lattice parameters. Note that no superposed periodicity is observed, which proves the single surface observation capability.

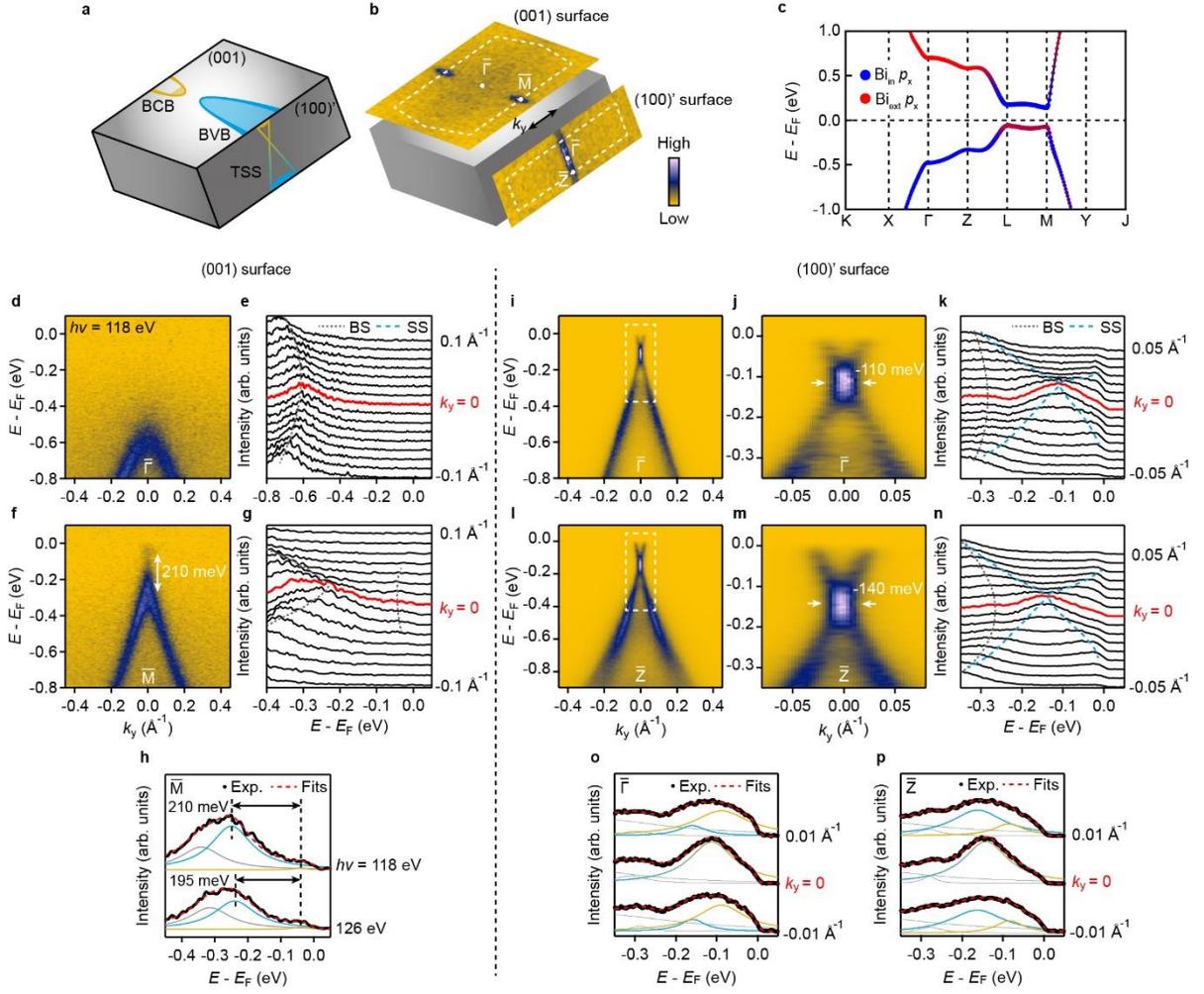

**Fig. 2 Surface-selective topological characterizations of BIB**

**a,** A schematic overview for the WTI nature of BIB based on observations that include two surface Dirac crossings on the (100)' surface and a trivial bulk gap on the (001) surface. TSS, BVB, BCB stand for topological surface states, bulk valence band, and bulk conduction band, respectively. **b,** FSs measured on the (001) and (100)' surfaces. **c,** First-principles bulk calculations for BIB. The valence and conduction bands near the Fermi level, mainly from the two $p_x$ orbitals (blue and red) of the interior and exterior Bi atoms ($Bi_{in}$ and $Bi_{ex}$ in Fig. 1a), form a global bulk band gap in 3D BZ, with two band inversions at the L and M points. **d,e,** Band dispersions across $\bar{\Gamma}$ on the (001)-projected surface BZ and the corresponding EDCs between $k_y$ = -0.1 Å$^{-1}$ to 0.1 Å$^{-1}$. Only the bulk valence band (grey dotted line) is observed whereas surface states are absent. **f,g,** Band dispersions across $\bar{M}$ on the (001)-projected surface BZ and the corresponding EDCs. **h,** EDCs at $\bar{M}$ measured with $hv$ = 118 eV and 126 eV (black dots). Each is fitted with three Lorentzian functions convoluted with the Fermi-Dirac distribution and Gaussian experimental resolution. Orange and cyan Lorentzians correspond to the bulk conduction and valence bands, respectively. **i**, Band dispersions across $\bar{\Gamma}$ on the (100)'-projected surface BZ. **j,** Zoomed-in view of the white box in **i**, showing the Dirac crossing at -110 meV. **k,** EDCs between $k_y$ = -0.05 Å$^{-1}$ and 0.05 Å$^{-1}$ from **j**. **l-n,** Equivalent plots as **i-k** for the $\bar{Z}$ point on the (100)'-projected surface BZ with a Dirac crossing at -140 meV. **o,p**, Fitting of the EDCs near $\bar{\Gamma}$ and $\bar{Z}$. Two grey Lorentzians at -0.3 eV are from bulk states. Orange, cyan, and gradient-colored (only for cuts at $k_y$ = 0) Lorentzian curves close to $E_F$ are from surface states. Note that single peak fitting works best to reproduce our experimental spectra at $k_y$ = 0 for both $\bar{\Gamma}$ and $\bar{Z}$, characterizing gapless nature of the TSS.

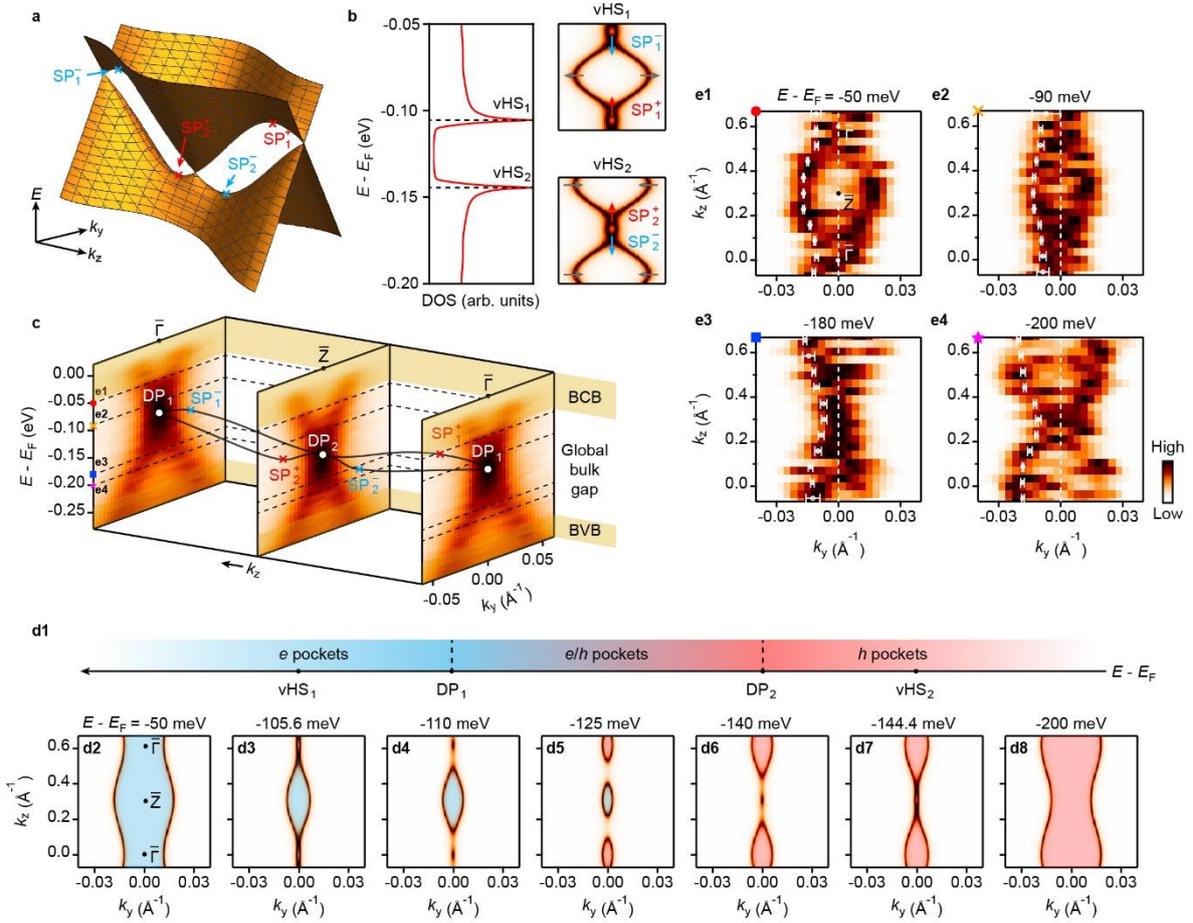

**Fig. 3 Saddle points and Lifshitz transitions of WTI topological surface states**

**a,** 3D electronic structure given by the minimal model Eq. (1). Two pairs of helical saddle points are labeled as $SP_{1,2}^{\pm}$. Subscripts 1 (2) indicate the energy location higher (lower) than that of the Dirac points, and superscripts $\pm$ indicate the spin polarizations. **b**, Left: Calculated surface state density of states using the minimal model Eq. (1). Two saddle-point van Hove singularities (vHS$_{1,2}$) are shown, corresponding to the energy locations of $SP_{1,2}^{\pm}$. Right: Spin textures on CESs at vHS$_{1,2}$ are also shown. **c**, 3D plot for a comprehensive description of the electronic structure on the (100)' surface. DP$_{1,2}$ show the Dirac points at $\bar{\Gamma}$ and $\bar{Z}$, respectively. $SP_{1,2}^{\pm}$ are also annotated. BVB and BCB refer to bulk valence and conduction bands respectively determined from Fig. 2g. **d**, CESs obtained from the minimal model Eq. (1) at key energies indicated by the bar along the energy axis showing locations for the van Hove singularities and Dirac points energies. Shades on CESs indicate electron- (blue) and hole- (red) pockets. **e**, Experimentally obtained CESs above and below vHS$_{1,2}$ and DP$_{1,2}$, demonstrating the contraction and expansion of the pockets at $\bar{\Gamma}$ and $\bar{Z}$ with decreasing energy as shown in **d**. Colored marks on the left top corners indicate the energy positions in **c**. White dots with error bars (standard deviation for the results) are fitted positions of CESs, which are used for quantitative determination of $v_y$ and $v_z$ in the minimal model Eq. (1).

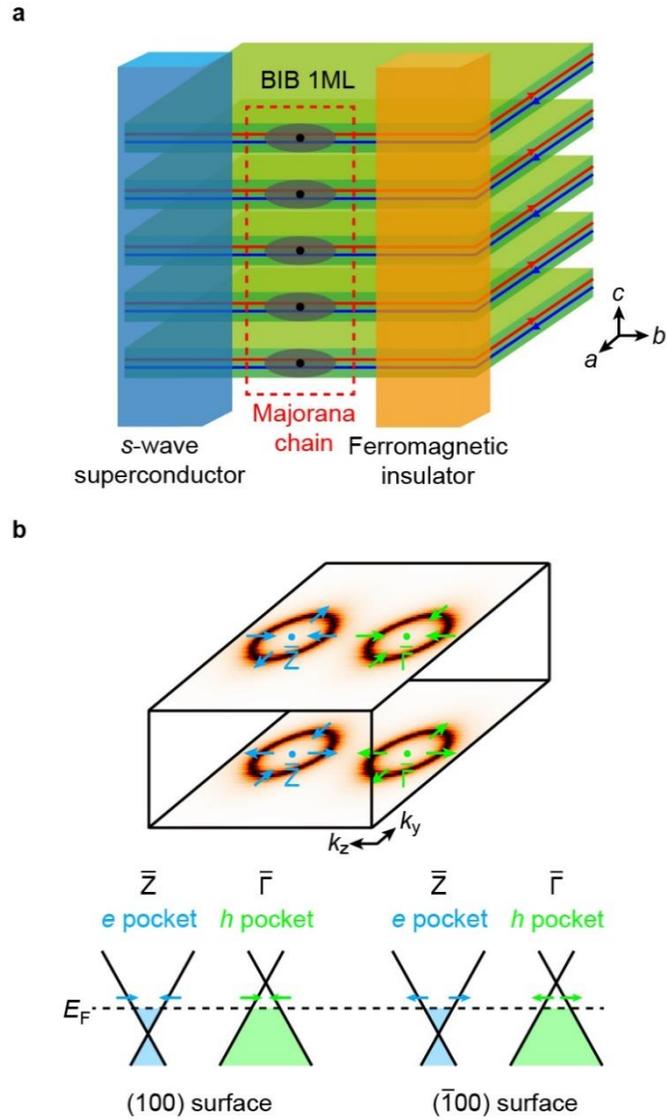

**Fig. 4 Possible topological many-body physics using BIB**

**a,** Emergence of Majorana chain when the (100)' TSS is proximitized to a linear junction of a ferromagnetic insulator and an *s*-wave superconductor. **b**, Unconventional exciton condensation by perfect nesting condition between electron and hole pockets in the TSS where their spin textures are only half-aligned for both the intra- and inter-surface pairing channels.